\documentclass[twocolumn,twoside,slac_two]{revtex4}
\usepackage{graphicx}
\usepackage{fancyhdr}
\newcommand{\ifb}{\mbox{fb$^{-1}$}}%  Inverse femtobarns.
\newcommand{\ipb}{\mbox{pb$^{-1}$}}%  Inverse picobarns.
\newcommand{\TeV}{\ifmmode {\mathrm{\ Te\kern -0.1em V}}\else
  \textrm{Te\kern -0.1em V}\fi}%
\newcommand{\GeV}{\ifmmode {\mathrm{\ Ge\kern -0.1em V}}\else
  \textrm{Ge\kern -0.1em V}\fi}%

\newcommand{\ttbar}{\ensuremath{t\bar{t}}}

\def\pt{\ensuremath{p_T} }
\def\met{\mbox{\ensuremath{\, \slash\kern-.6emE_{T}}}}
\newcommand\dd{{\rm d}}

\begin{document}

\title{Top quark pair cross section prospects in ATLAS}


\author{Andrei Gaponenko}
\affiliation{Lawrence Berkeley National Lab, Berkeley, CA 94720, USA}

\begin{abstract}
The observation of the top quark will be an important milestone in
ATLAS.  This talk reviews methods that ATLAS plans to use to observe
the top quark pair production process and measure its cross section.
\end{abstract}

\maketitle

\thispagestyle{fancy}


%%%%%%%%%%%%%%%%%%%%%%%%%%%%%%%%%%
\section{Introduction}
The motivations to study top quarks are manifold.  The top quark can
be used as a probe for new physics.  Top pair production process
presents a dominant background to many New Physics searches.
% ----------------------------------------------------------------
%(ATL-PHYS-PUB-2009-084) 
% ----------------------------------------------------------------
Finally, it will be a valuable tool for detector 
calibration at the LHC.  Establishing a top quark signal will be an
important milestone in the ATLAS physics program.

The results in this talk are based upon simulations with 14\TeV{}
center of mass energy.  Unless stated otherwise, the studies presented
here are documented in~\cite{csc-book}.

In hadron collisions, the top quark can be produced via either the
strong (\ttbar{}) or electroweak (single top) process.  While the pair
production was established at the Tevatron in 1995
\cite{top-discovery-cdf}, \cite{top-discovery-d0}, it took until 2009
to observe the electroweak process \cite{tevatron-single-top-d0},
\cite{tevatron-single-top-cdf}.  While at the LHC we expect to observe
single top in a much shorter time frame, the initial focus will be on
the \ttbar{} process.

The production cross section of top quark pairs increases from about
7~pb at the Tevatron to about 900~pb ($\sqrt{s}=14\TeV$) or 400~pb
($\sqrt{s}=10\TeV$) at the LHC.  For example, with 200~\ipb{} of
integrated luminosity at 10\TeV{} about 80k \ttbar{} would be
produced.  Another important consideration is that the production
cross section of the dominant background, $W+{}$jets, increases more slowly
with energy than the signal cross section.  That improves $S/B$ and
makes the \ttbar{} process easier to observe than at the Tevatron.

According to the Standard Model, the top quark decays almost
exclusively into a $b$ quark and a $W$ boson.  Top quark pair decays
are classified according to the decay modes of the two $W$ bosons into
all hadronic, lepton+jets, and dilepton channels.  In ATLAS the
initial focus will be on decays that have either an electron or a muon
from $W$ decay in the final state.  Such decays will be an important
detector validation and calibration tool because they contain energetic jets,
leptons, b-jets, and large missing transverse energy, allowing 
ATLAS subdetectors to be exercised.   Jet energy scale and b-jet
identification efficiency can be calibrated using \ttbar{} events.

The lepton+jets channel has a relatively large branching fraction
(about 15\% per flavor), and full kinematic information for the
hadronic top quark is preserved, allowing for a mass peak
reconstruction.  The background in this channel may be relatively
large, but it is dominated by one process, $W+{}$jets.

The dilepton channel has a much lower background.  An interesting
feature of the $e\mu$ channel is that one can measure trigger
efficiencies directly on \ttbar{} events.  But dilepton channels also
have a much smaller branching fractions (about 1\% for $\mu\mu$, $ee$
or 2\% for $e\mu{}$), and more kinematic information is lost because
of the two escaping neutrinos.

% add a paragraph about trigger?

%%%%%%%%%%%%%%%%%%%%%%%%%%%%%%%%%%
\section{Total cross section measurements}

\subsection{Lepton+jets analysis}

\paragraph*{Event preselection}
\begin{itemize}
\item Lepton $\pt>20\GeV$, $|\eta|<2.5$
\item $\met>20\GeV$
\item ${}\ge4$ jets $\pt>20\GeV$, $|\eta|<2.5$
\item Including ${}\ge3$ jets $\pt>40\GeV$
\end{itemize}
Note that no b-tagging is used.  We require either a ``medium''
electron, or a ``combined'' muon~\cite{csc-book}.

\begin{figure}[tbhp]
\centering
\includegraphics[width=80mm]{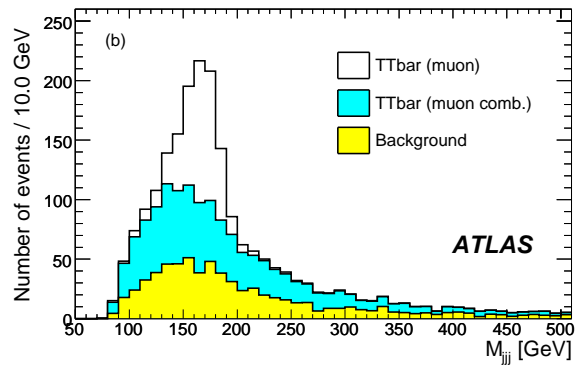}
\caption{A stack plot of the 3-jet invariant mass for \ttbar{} events
  in the muon+jets channel
  that were reconstructed correctly (clear histogram), where a wrong
  jet combination was chosen (cyan histogram), and non \ttbar{} 
  background (yellow
  histogram at the bottom).  The distributions are normalized to
  100~\ipb{} at 14~\TeV.\label{fig:total_mjjj}}
\end{figure}

There are different methods to reconstruct the hadronically decaying
top quark, that is, to determine which of the jets in the events are
produced by decay products of that quark, in a \ttbar{} event.  The
method used here is based on the fact that at the LHC top quarks are
not produced at rest but have a non-zero $\pt$.  Since QCD has no
intrinsic mass scale, the average $\pt$ of an object is proportional
to the mass of the object.  The top quark is the heaviest Standard
Model object, so we select the 3 jets that maximize
$\sum_{i=1}^{3}{\vec{p}}_{T,i}$ as the daughters of the top, and treat
the invariant mass of the three jet combination as the mass of the
reconstructed top candidate.  Figure~\ref{fig:total_mjjj} shows
invariant mass spectra for correctly and incorrectly reconstructed
signal events, and also the background contribution (dominated by $W+{}$jets).

After a 3-jet combination is assigned to a top quark, we apply an
additional cut:  at least one of the pairs of jets originating from
the top quark should have an invariant mass compatible with that of
the $W$:
\begin{itemize}
\item $|M_{jj}-M_{W}|<10\GeV$ \\
  for a hadronic-top jet-pairing
\end{itemize}  
The expected numbers of events in 100~\ipb{} at 14~\TeV{} for the signal
and main backgrounds after the selection cuts are summarized in
Table~\ref{table:total_counts}.

\begin{table}[tbhp]
\begin{center}
\caption{Expected event counts in the lepton+jets channel for
  100~\ipb{} at 14~\TeV{}.}
\begin{tabular}{|l|r|r|}
\hline
 & $e$ & $\mu$ \\\hline
\vrule width0pt height2.2ex depth0.8ex
signal \ttbar & 1262 & 1606 \\
\hline
$W$+jets & 241 & 319 \\
single top & 67 & 99\\
$Z$+jets & 35 &23 \\
other non-QCD  & 31 & 54\\
\hline
Total background  & 374 & 495\\
\hline
S/B & 3.4 & 3.2\\
\hline
\end{tabular}
\label{table:total_counts}
\end{center}
\end{table}

\begin{figure}[tbhp]
\centering
\includegraphics[width=80mm]{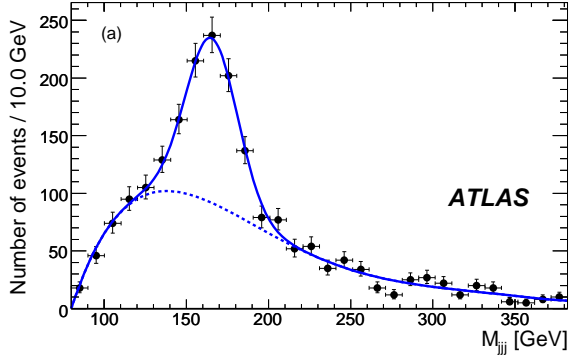}
\caption{An example of the invariant mass fit in the muon+jets
  channel. Normalized to
  100~\ipb{} at 14~\TeV.\label{fig:total_fit}}
\end{figure}

The simplest (``counting'') method to extract the cross section will use the number
of data events in the signal region and subtract background estimated
from simulations.  Another possibility is to fit the 3-jet invariant
mass distribution to estimate the number of signal events while
normalizing background to data.  An example of a fit is shown in
Figure~\ref{fig:total_fit}. 

% Both with W mass constraint

\begin{table}[htbp]
\begin{center}
\caption{Relative uncertainties (in percent) of a \ttbar{} cross
  section measurement for the counting and the fitting methods in the
  electron+jets channel.}
\begin{tabular}{|l|c|c|}
\hline
Source                & Counting & Fitting \\
\hline                                     
Statistical           & 3.5      & 10.5    \\
50\% more $W+{}$jets  & 9.5      & 1.0     \\
Jet energy scale      & 9.7      & 2.3     \\
ISR/FSR               & 8.9      & 8.9     \\
PDFs                  & 2.5      & 2.5     \\
Lepton ID             & 1.0      & 1.0     \\
Trigger eff           & 1.0      & 1.0     \\
Shape of fit func     &   -      & 14.0    \\
\hline
\end{tabular}
\label{table:total_sys}
\end{center}
\end{table}

A summary of statistical and dominant systematic uncertainties for
these two methods is shown in Table~\ref{table:total_sys}.  One can
see that the two methods are complementary because their leading
systematics are different: the counting method is more sensitive to
the $W+{}$jets background uncertainty and the jet energy scale, but
the fitting method has a large systematic associated with the shape of
the fitting function.  The expected final uncertainties of a \ttbar{}
cross section measurement using 100~\ipb{} of data at 14~\TeV, using a
combination of the electron+jets and muon+jets channels, are the
following.  Counting method:\\
$$\Delta\sigma/\sigma =
({}\pm3\text{(stat)}\pm16\text{(sys)}\pm3\text{(pdf)}\pm5\text{(lumi)})\%$$
Fitting method:
$$\Delta\sigma/\sigma =
({}\pm7\text{(stat)}\pm15\text{(sys)}\pm3\text{(pdf)}\pm5\text{(lumi)})\%$$

\subsection{Dilepton channel}

\begin{figure*}[tbp]
\centering
\begin{tabular}{ccc}
 \includegraphics[width=50mm]{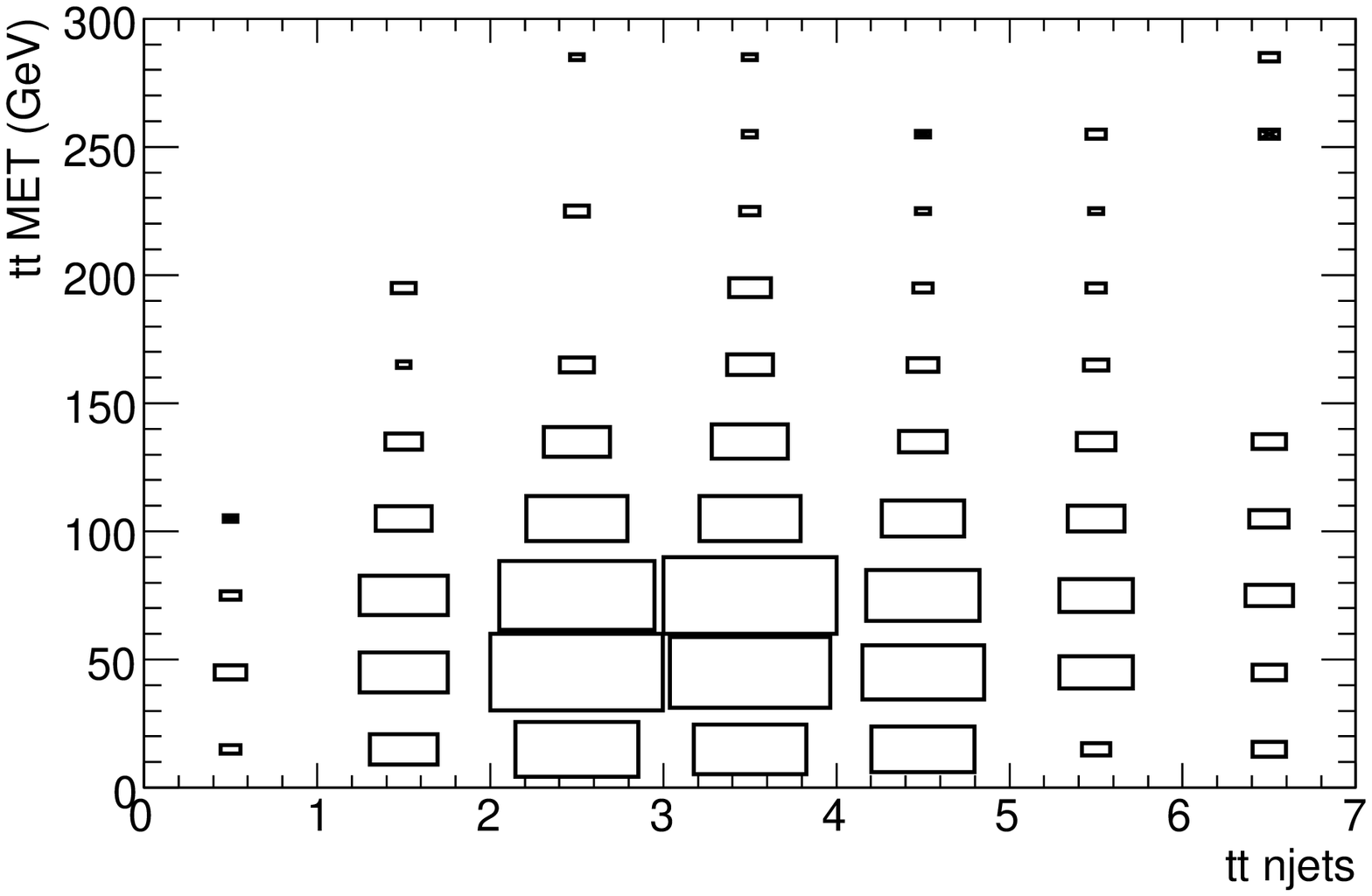}
%&\includegraphics[width=50mm]{../eps/ww_nj_met.eps}
&\includegraphics[width=50mm]{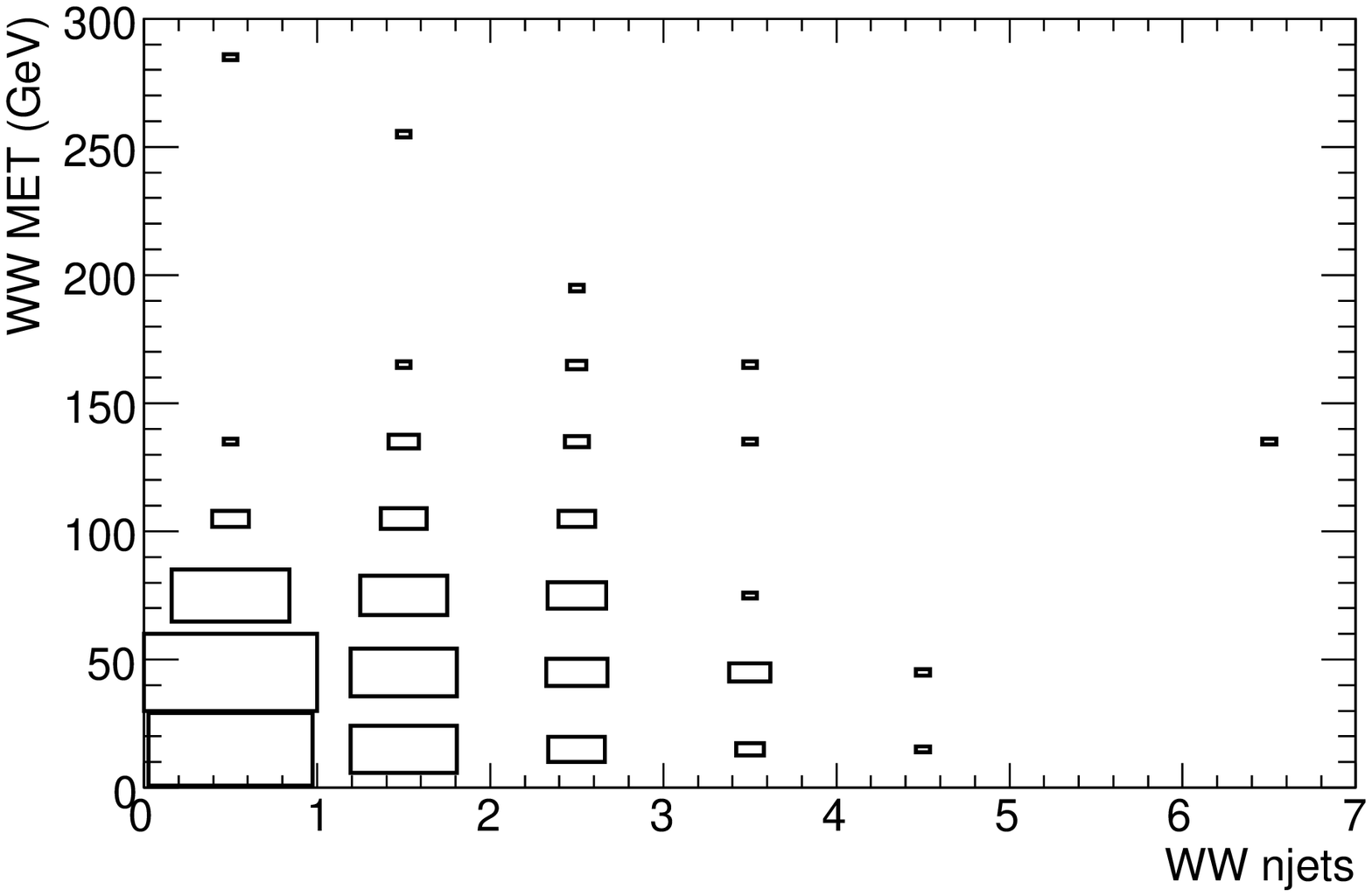}
%&\includegraphics[width=50mm]{../eps/ztt_nj_met.eps}
&\includegraphics[width=50mm]{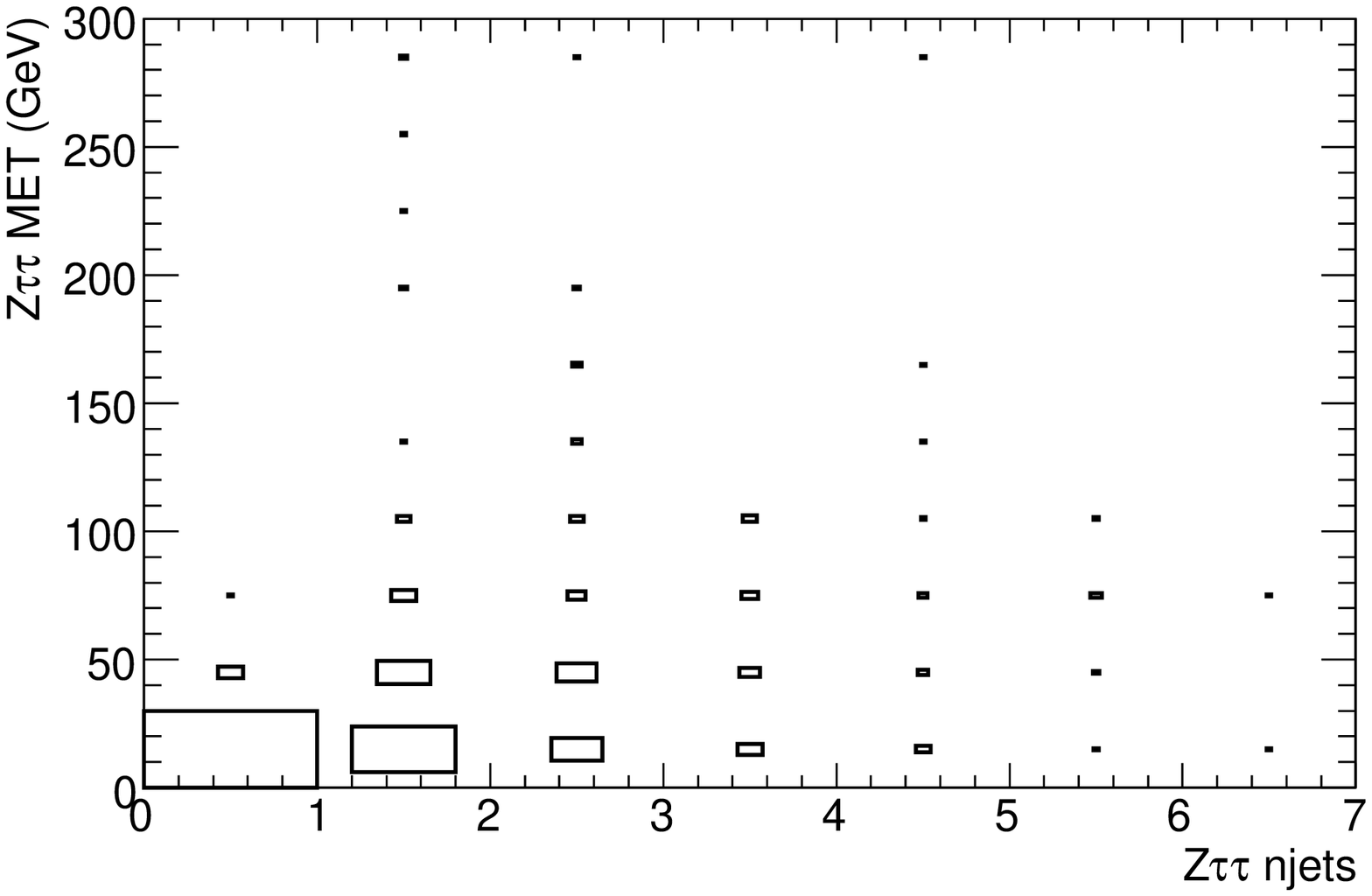}
\end{tabular}
\caption{Example of $\met$ vs number of jets distributions used in by
  the template fit method for \ttbar{} (left), $WW$ (middle), and
  $Z\to\tau\tau$ (right) processes.  This illustration uses the
  $e\mu$ channel.\label{fig:total_templates}}
\end{figure*}

ATLAS has developed several methods to measure the \ttbar{} production
cross section in the dilepton (lepton=$e$, $\mu$) channel.  The method
utilizing a template fit is presented here.

\begin{figure}[tbhp]
\centering
\includegraphics[width=80mm]{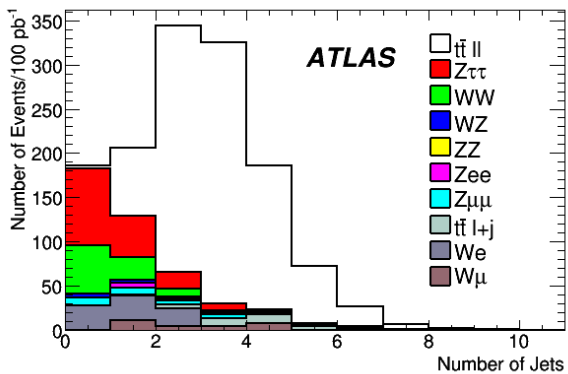}
\caption{Total composition of the inclusive di-lepton selection versus
  number of jets. Normalized to 100~\ipb{} at
  14~\TeV.\label{fig:total_template_fit}}
\end{figure}

\paragraph*{Event selection}
\begin{itemize}
\item 2 opposite sign leptons ($ee$, $e\mu$, $\mu\mu$)
  with \\ $\pt>20\GeV$,  $|\eta|<2.5$.

\item For the $\mu\mu$ channel: require that the $\met$ vector is not
  parallel or anti-parallel to a muon pt vector.

\item For same flavor channels: veto $Z$ mass region, and require
  $\met>35\GeV$.  There is no $\met$ cut for the $e\mu$ chanel.
\end{itemize}
Here we use ``tight'' electrons, and require, $dR(\mu,\text{jet})>0.2$.
Event passing those cuts are filled into 2-dimensional histograms of
$\met$ vs the number of jets, such as shown in
Figure~\ref{fig:total_templates}.  The distributions of the \ttbar{}
signal and different backgrounds in these two variables have different
shape, thus allowing a data distribution to be fit with a linear
combination of simulated templates to extract the number of signal
events. 
A combination of the templates according to their Standard Model cross
sections, projected on the jet multiplicity axis, 
is shown in Figure~\ref{fig:total_template_fit}.  
The expected uncertainty of the measurement assuming 100~\ipb{} of
data at 14~\TeV{} is
$$\Delta{\sigma}/\sigma =
(4\text{(stat)}\pm4\text{(sys)}\pm2\text{(pdf)}\pm5\text{(lumi)})\%.
$$

%%%%%%%%%%%%%%%%%%%%%%%%%%%%%%%%%%
\section{Differential cross section}

Measuring $\sigma_{\ttbar}$ vs kinematic variables of the \ttbar{}
system allows one to:
\begin{itemize}
  \item test QCD predictions
  \item validate our understanding of detector acceptance
  \item provide a way to search for New Physics
\end{itemize}
One of manifestations of physics beyond the Standard Model may be a
heavy resonance decaying into a \ttbar{} pair.  It could be observed
in a $\dd\sigma_{\ttbar}/\dd{}m_{\ttbar}$ measurement, which we will
discuss in this section.

\begin{figure}[tbhp]
\centering
\includegraphics[width=80mm]{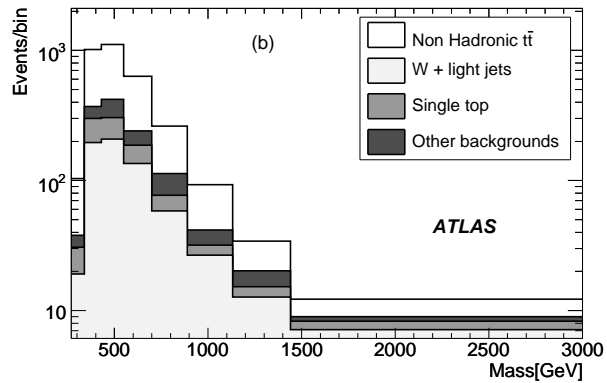}
\caption{Invariant mass distribution of the \ttbar{} system using
  conventional event reconstruction in the lepton+jets
  channel. Normalized to 100~\ipb{} at
  14~\TeV.\label{fig:diff_xsec_conventional}}
\end{figure}

\begin{figure}[tbhp]
\centering
\includegraphics[width=80mm]{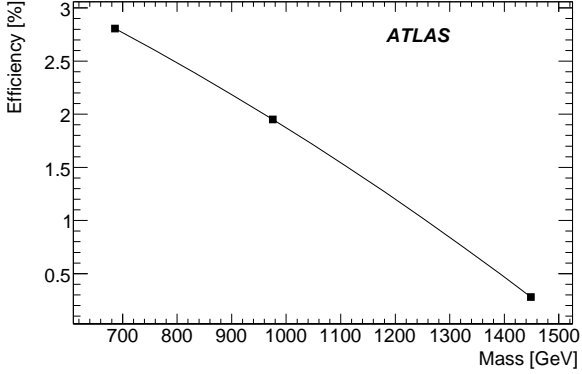}
\caption{Efficiency of a conventional \ttbar{} reconstruction method 
vs the invariant mass of the \ttbar{} system.\label{fig:diff_conventional_eff}}
\end{figure}

A straightforward way to measure the differential cross section is to
extend the lepton+jets analysis presented above, and perform on
selected events a kinematic fit of the 4 leading jets, lepton, and
$\met$, using $m_{top}$ and $m_W$ constraints to reconstruct the
invariant mass of the \ttbar{} system.  The expected shape of the
\ttbar{} invariant mass distribution is shown in
Figure~\ref{fig:diff_xsec_conventional}.
% ----------------------------------------------------------------
% ----------------------------------------------------------------
However conventional methods of top reconstruction that require the decay
products of each top quark to be identified in the detector as
distinct jets or isolated leptons become inefficient for highly
boosted top quarks, as illustrated in Figure~\ref{fig:diff_conventional_eff}.
% ----------------------------------------------------------------
% ----------------------------------------------------------------
Therefore such methods are not well suited for heavy resonance
searches.  A new method for top pair reconstruction targeted for heavy
resonance searches in the lepton+jets channel, based
on \cite{Thaler:2008ju}, \cite{Butterworth:2002tt}, is suggested
in \cite{ATL-PHYS-PUB-2009-081}.  The ATLAS note uses a generic narrow
$Z'$ decaying to \ttbar{} as a benchmark.

\begin{figure*}[tbp]
\centering
\begin{tabular}{ccc}
 \includegraphics[width=0.5\textwidth]{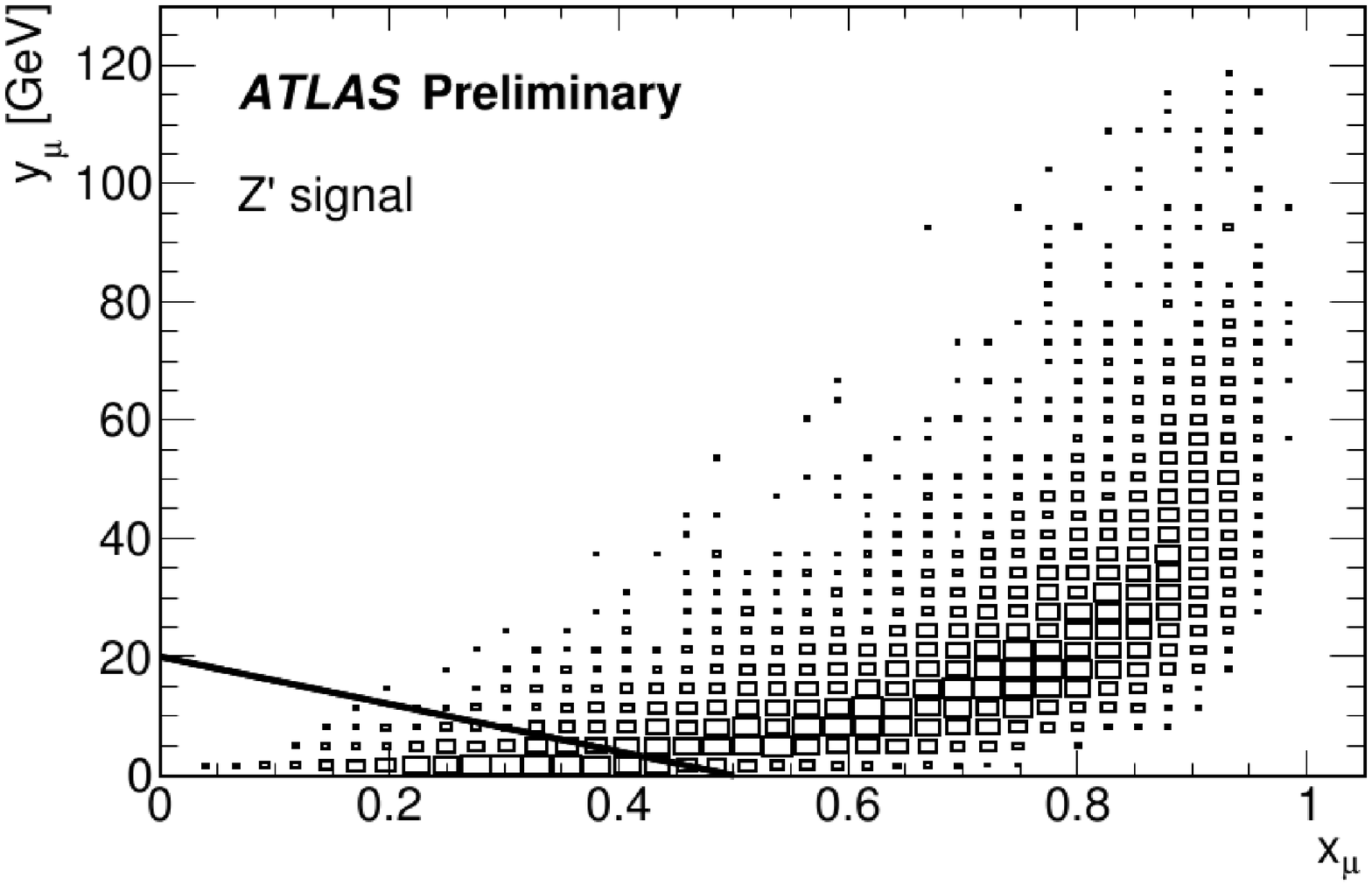}
%&\includegraphics[width=0.5\textwidth]{../eps/ATL-PHYS-PUB-2009-081.fig2_b.eps}
&\includegraphics[width=0.5\textwidth]{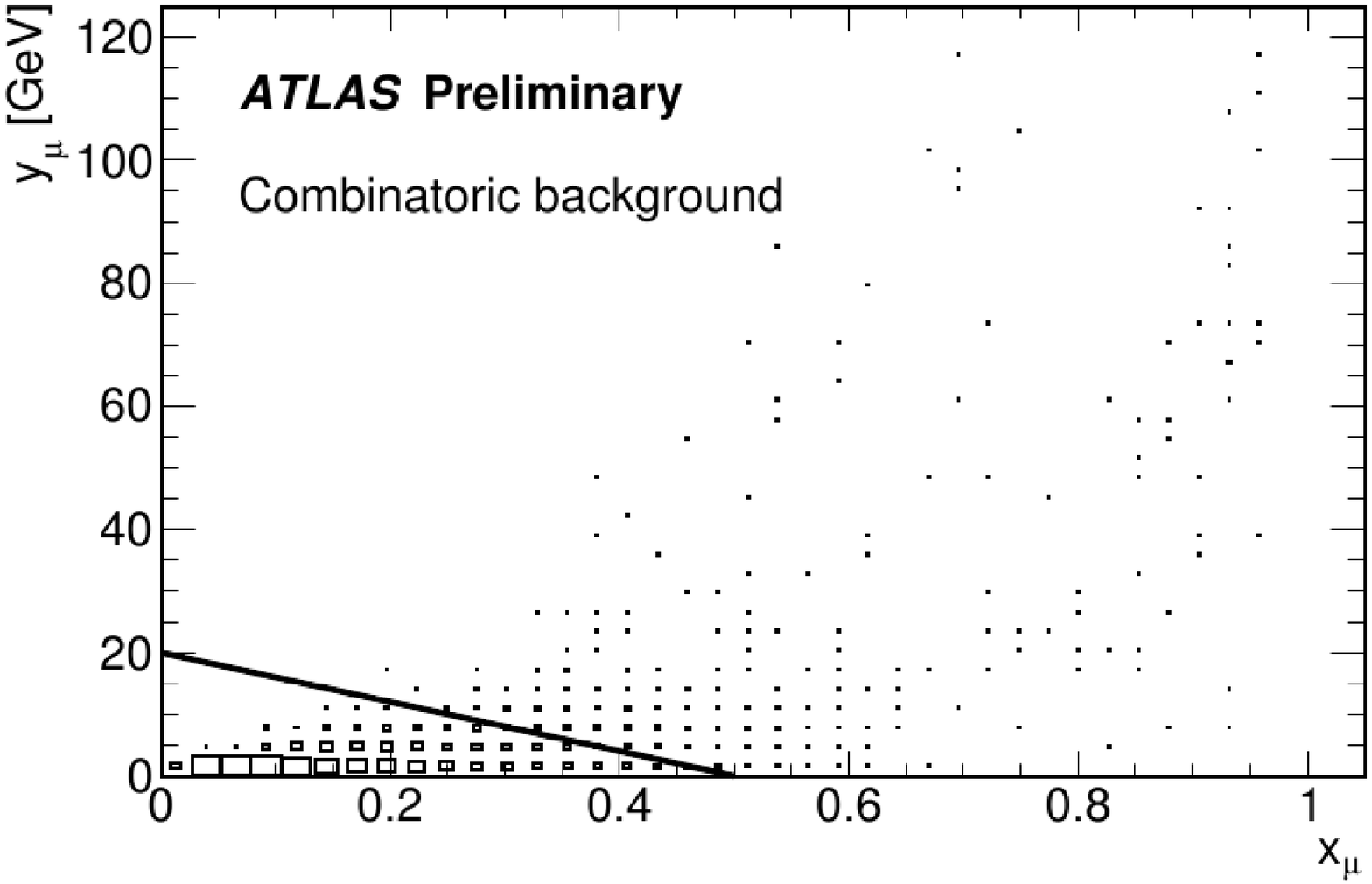}
\end{tabular}
\caption{2D distributions of $(x_{\ell}, y_{\mu}$ for correct (left),
  and wrong (right) combinations of a muon and a jet in muon+jets
  $Z'\to\ttbar$ decays. Events are required to lie above the
  line to pass selection cuts.\label{fig:diff_leptonic_cut}}
\end{figure*}

\paragraph*{Leptonic top reconstruction}
In the muon+jets channel, the reconstruction starts by identifying the
hardest muon in the event with $\pt>20\GeV$ and $|\eta|<2.5$, without
making an isolation requirement.  A jet with $\pt>200\GeV$ within
$\Delta{R(\mu,\text{jet})}<0.6$ of the selected muon is assumed to
originate from the same top quark as the lepton.  To suppress wrong
$\mu+{}$jet combinations, a cut on the following two variables is
used:
$$x_{\mu} = 1-m^2_{\text{jet}}/m^2_{\text{jet}+\mu}$$ where
$m_{\text{jet}}$ is the invariant mass of the jet and
$m_{\text{jet}+\mu}$ is the invariant mass of the muon+jet candidate
system, and
$$y_{\mu} = p_{\mu\perp{}\text{jet}}\times\Delta{R(\mu,\text{jet})}$$
where $p_{\mu\perp{}\text{jet}}$ is the transverse momentum of the
muon with respect to the jet.  The cut, 
$$
y_{\mu} > (-40x_{\mu} +20\GeV),
$$
is illustrated in Figure~\ref{fig:diff_leptonic_cut}.  Reconstruction
in the electron+jets channel is similar, but the jet momentum has to
be corrected to avoid double counting of the electron energy, and
additional cuts on the jet to suppress QCD background are applied.

All missing transverse energy of the event is assumed to come from the
neutrino.  A requirement of 
$$
\met>20\GeV
$$ is imposed, and a $W$ mass constraint is applied to deduce the $z$
component of the neutrino momentum.  The smaller $|p^{\nu}_z|$ is
chosen in case of ambiguity, and the real part is used in case only
complex solutions exist.  A further cut of 
$$
\Delta{R}(\nu,l)<1
$$
is imposed to reject misreconstructed neutrinos, as the decay products
are expected to be collinear.
Finally, the four momentum of the leptonic top is computed as
$$p_{\text{~top},\ell}=p_{\ell}+p_{\text{jet}}+p_{\nu}.$$ 

\paragraph*{Hadronic top reconstruction}
After a leptonic top is reconstructed, the hardest of the remaining
jets (which are $k_{\perp}$ jets with $D=0.6$) is considered a
hadronic top candidate.  It must satisfy
$$
\pt>300\GeV. 
$$ The fine granularity of the ATLAS calorimeter allows jet
substructure to be resolved and used to suppress backgrounds.  A likelihood
variable is build from jet mass and jet substructure variables
($k_{\perp}$ splitting scales), and a cut on this variable is applied
to separate top jets from those coming from other sources.
 
% ----------------------------------------------------------------
\begin{figure}[tbhp]
\centering
\includegraphics[width=80mm]{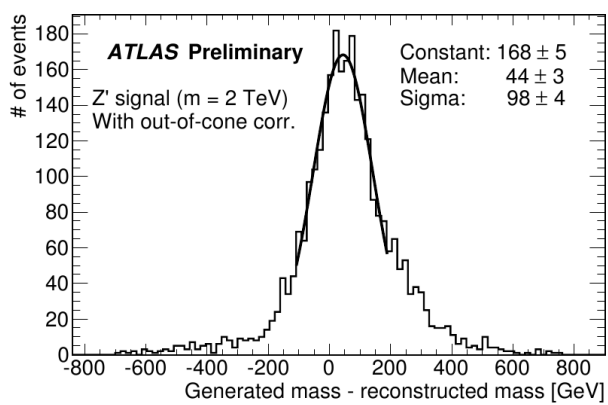}
\caption{Generated minus reconstructed mass of a $2\TeV$
  $Z'$.\label{fig:diff_zprime}}
\end{figure}
% ----------------------------------------------------------------

After applying an out of cone energy correction to the reconstructed
tops, the $Z'$ invariant mass is computed as
$$m_{Z',\text{rec}}^2=(p_{\text{top},\ell}+p_{\text{top,had}})^2.$$
The resulting distribution for a 2~\TeV{} $Z'$ is shown in
Figure~\ref{fig:diff_zprime}.

A signal region of $\pm1\sigma_{\text{reco}}$ around the $Z'$ peak is
defined.  The signal reconstruction efficiency is $0.094\pm0.002$, and
estimated background counts in that region 
are summarized in Table~\ref{table:diff_zprime_bg}.
The systematic uncertainty on Standard Model \ttbar{} background
includes 10\% from the total cross section, and 15\% due to
extrapolation from a restricted region of the phase space. (Standard
Model \ttbar{} samples for this study 
were generated only at high \ttbar{} invariant
masses to get sufficient statistics.)

\begin{table}[htbp]
\begin{center}
\caption{Estimated background counts in the signal region for a 2\TeV{}
  $Z'$ with 1\ifb{} at 14\TeV.}
\begin{tabular}{l|l}
%\hline & $m_{Z'}=2\TeV$ &  $m_{Z'}=3\TeV$ \\
\hline
\vrule depth 0pt height2ex width0pt
SM $\ttbar$ & $21.9\pm1.0\pm3.9\text{(sys)}$ \\
QCD multijet & $\hphantom{2}1.9\pm0.5$\\
\hline
Total & $23.8\pm4.1$ 
\end{tabular}
\label{table:diff_zprime_bg}
\end{center}
\end{table}

A Bayesian technique was used to include uncertainties on the
background, signal acceptance and luminosity in an estimate of the
sensitivity to a narrow resonance decaying to \ttbar{}.  The expected
95\% CL on $\sigma\times{}Br(\ttbar)$ with 1\ifb{} at 14\TeV{} are
\begin{itemize}
\item {550\ fb for $M_{Z'}=2\TeV$}
\item {160\ fb for $M_{Z'}=3\TeV{}$}
\end{itemize}

%%%%%%%%%%%%%%%%%%%%%%%%%%%%%%%%%%
\section{Summary}

ATLAS will have a rich top physics program.  In addition, top pair events will
provide valuable tool for detector validation and calibration.
Before exploring and exploiting properties of the top quark we need to
observe it.  A cross section measurement at a new center of mass
energy is also an important physics result.

This talk presented some of the methods that ATLAS will use to measure
\ttbar{} cross section in the lepton+jets and dilepton channels.   It
also outlined a new method of reconstructing \ttbar{} events in the
lepton+jets channel that is targeted to high mass resonances decaying
into \ttbar{}, and showed the expected sensitivity of the new method.

Establishing top signal is an important milestone in the ATLAS physics
program, and ATLAS is prepared to achieve this over a wide range of LHC
$\sqrt{s}$, integrated luminosity of usable data, and detector
performance.

\bigskip % extra skip inserted
\begin{acknowledgments}
  This work was supported by the Director, Office of Science,
  Office of High Energy Physics, of the U.S. Department of Energy
  under Contract No. DE-AC02-05CH11231.
\end{acknowledgments}

\bigskip % extra skip inserted

\end{document}